\documentclass[%
 reprint,
superscriptaddress,
 amsmath, amssymb,
 aip,
 apl,
]{revtex4-2}
\usepackage{silence}
\usepackage{graphicx}%
\usepackage{dcolumn}%
\usepackage{bm}%

\usepackage[utf8]{inputenc}
\usepackage[T1]{fontenc}
\usepackage{mathptmx}
\usepackage{etoolbox}
\usepackage[dvipsnames]{xcolor}
\usepackage{siunitx}
\usepackage{float}
\usepackage{booktabs}
\usepackage{multirow}
\usepackage{orcidlink}

\begin{document}

\title{Robotic chip-scale nanofabrication for superior consistency}

\author{Felix M. Mayor\,\orcidlink{0000-0001-9849-3517}}
\email{fmayor@stanford.edu}
\author{Wenyan Guan\,\orcidlink{0009-0006-9284-7920}}
\author{Erik Szakiel\,\orcidlink{0009-0000-2318-1180}}
\author{Amir H. Safavi-Naeini\,\orcidlink{0000-0001-6176-1274}}
\author{Samuel Gyger\,\orcidlink{0000-0003-2080-9897}}
\email{gyger@stanford.edu}
\affiliation{Department of Applied Physics and Ginzton Laboratory, Stanford University, 348 Via Pueblo Mall, Stanford, California 94305, USA}

\date{\today}%

\begin{abstract}
Unlike the rigid, high-volume automation found in industry, academic research requires process flexibility that has historically relied on variable manual operations.
This hinders the fabrication of advanced, complex devices.
We propose to address this gap by automating these low-volume, high-stakes tasks using a robotic arm to improve process control and consistency.
As a proof of concept, we deploy this system for the resist development of Josephson junction devices. 
A statistical comparison of the process repeatability shows the robotic process achieves a resistance spread across chips close to \qty{2}{\percent}, a significant improvement over the $\sim \qty{7}{\percent}$ spread observed from human operators, validating robotics as a solution to eliminate operator-dependent variability and a path towards industrial-level consistency in a research setting.
\end{abstract}

\maketitle

Nanofabrication, the ability to engineer materials down to the nanometer scale, is a cornerstone of modern technology.\cite{Gatzen.Saile.ea:2015} 
Its biggest impact is semiconductor electronics, but it is now an integral component in almost every field of advanced science.
The technology is defined by two disparate worlds. 
On one hand, industrial foundries run billion-dollar factory floors, with highly optimized, almost completely automated processes inside cleanroom facilities on \SI{300}{\milli\meter} wafers to produce silicon electronics at massive scale.
On the other hand, in universities and research centers, cleanrooms operate as high-mix, low-volume environments. 
Here, flexibility, small-area source materials, and reconfigurable operations are the norm to explore new materials, techniques and physical properties to provide the foundation for future technology cycles.

Crucial for exploring new ideas, this artisanal research environment spans not only silicon-adjacent fields like 
integrated nanophotonics\cite{Yang.Jeon.ea:2025}, 
brain-inspired computing\cite{Mehonic.Ielmini.ea:2024}, and 
quantum-enhanced technology\cite{Laucht.Hohls.ea:2021}, 
but also biomedical applications\cite{Luttge:2016}, 
neuroscience\cite{hong2019novel},
and next-generation sensors\cite{Kar.Guerra.ea:2024}.
Such explorations are characterized by highly manual operations, small sample sizes, significant process flexibility, and a wide variety of material stacks.
While this flexibility drives scientific discovery, it also represents a critical bottleneck.
The reliance on humans for complex, multi-step processes introduces significant variability, limiting reproducibility and hindering the fabrication of highly complex integrated devices.

\begin{figure}[!tbp]
\includegraphics[width=\columnwidth]{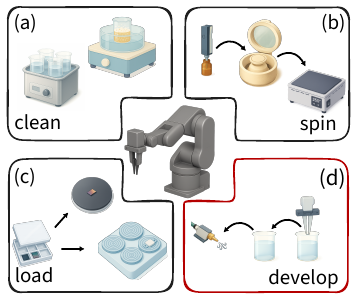}
\caption{\label{fig:concept}\textbf{Robotic Assisted Nanofabrication.} Robotic arms equipped with tweezers are the missing link for explorative nanofabrication at sample scale. 
We foresee four tasks enabled by picking up and handling samples that promise improvement in reliability and safety. 
\textbf{(a)}\,Cleaning samples in ultrasonic solvent baths or chemical cleans. 
\textbf{(b)}\,Spinning resist using electronic pipettes for volume control, followed by standard spinners and hotplates.
\textbf{(c)}\,Loading prepared substrates on carrier wafers for processing and using resist or oil for bonding. 
Accurate placement on tool carriers for lithography reduces loading time and increases tool utilization. 
\textbf{(d)}\,Developing samples by controlled liquid chemical exposure and final drying step. Explored in this work.
}
\end{figure}

Nanofabrication tools for lithography, deposition, and etching already offer high degrees of automation. 
The remaining gaps are in sample transfer, metrology, and most critically, in wet-chemical processing. 
These steps are often realized entirely by human action and are a major source of process uncertainty.

The recent developments of affordable, high-accuracy and table-top robotics present a solution to closing this gap for sample and small-batch fabrication.
While still a notable investment, their cost is comparatively negligible against the expenditure connected to a cleanroom operation.
Automating these manual steps paves the way towards a fully automated, flexible and high-fidelity nanofabrication pipeline for research and automated discovery.
We envision four core areas (see Fig.~\ref{fig:concept}) where such a robot will be immediately useful: (a) cleaning samples in solvents or aggressive chemicals, (b) preparing them for lithography, (c) accurate placement and fixation on sample carriers for etch, deposition or lithography, and (d) resist development.

The benefits range from improved chemical safety to higher and more repeatable throughput. 
To justify the investment, such a robot must offer more than just convenience: it must drive tangible improvements in process outcomes.
We select resist development as our case study because it offers significant potential for improving chip consistency.

We compare the fabrication of a Dolan-bridge style Josephson Junction (JJ)\cite{Niemeyer.Kose:1976, Dolan:1977}, a key building block for superconducting quantum computing and electronics\cite{GoogleQuantumAI:2025, AWSQuantumComputing:2025, Krantz.Kjaergaard.ea:2019}, specifically examining the difference between manual and robotic resist development.
We show a clear reduction in sample-to-sample variability in the robotic case by eliminating inter-operator variance, a common source of inconsistency in research environments where multiple researchers conduct the same process.

\begin{figure*}[!htbp]
\includegraphics[scale=1]{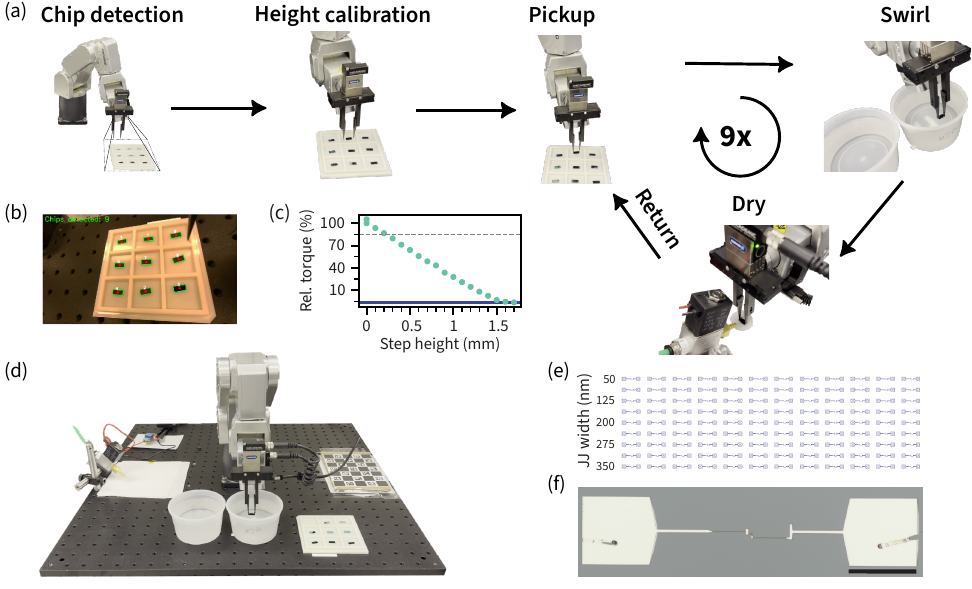}
\caption{\label{fig:fab} \textbf{Robotic Resist Development.} 
\textbf{(a)}\,After chip detection and hand-eye transformation, the tool-height is calibrated using torque-based feedback. 
Each chip is then individually picked up, swirled inside a developer solution and a rinse solution for \qty{40}{\second} and \qty{10}{\second} respectively, before being blow-dried for \qty{10}{\second} in a nitrogen flow. 
Afterwards the chip is returned to the chip-carrier. 
\textbf{(b)}\,OpenCV-based chip-edge detection allows free placement of chip-carrier by a human operator. 
\textbf{(c)}\,Reliable height detection for chip pick up. Relative torque on joint $5$ of the arm versus vertical step height, starting from an overdriven position (verified by an initial drop in torque of \qty{15}{\percent} upon lifting the tweezers, see dashed line). The arm raises the tweezers in \qty{0.1}{\milli\meter} steps until the torque change between steps drops below \qty{0.8}{\percent} (blue line), inferring that the tips have lifted off the chip box. This point establishes the reference height for grasping.
\textbf{(d)}\,Mecademic M500 Robot on an optical breadboard with access to the required fabrication steps for sample based wet-bench tasks.
\textbf{(e)}\,CAD of the chip showing the array of JJ test devices. In each row, there are $12$ identical devices, and we sweep the JJ width over the $9$ rows. 
\textbf{(f)}\,Microscope image of a fabricated JJ test device after probing. Scale bar: \qty{100}{\micro\meter}.
}
\end{figure*}

The detailed robotic developing process is described in Fig.~\ref{fig:fab}(a). A sample box with a $3\times3$ grid (Entegris H44-999-1415) is placed on a breadboard within reach of a robotic arm (Meca500-R4) with $6$ degrees of freedom and controlled via ROS2. The arm is equipped with a gripper (MEGP-25 LS) toolhead to which we attach a pair of tweezer tips (3D printed adapter to Techni-Pro 758TW0305 replacement tips). A camera (Realsense D405), mounted to the arm (link 5), points at the sample box. We run a chip detection algorithm based on OpenCV to extract the position and angle of all the chips placed in the sample box, see Fig.~\ref{fig:fab}(b). Previously performed hand-eye calibration allows us to transform coordinates from the camera to the tweezer tip frame (see \ref{SI:hand}). We use the in-plane coordinates to move the robotic arm to the first chip and hover the tweezers in the open position over the chip. Since the chip is only $\qty{0.5}{\milli\meter}$ thick, we cannot rely on the camera image for finding the correct height. 
Instead, we rely on a torque-based method to measure the grasping height. 
The robotic arm lowers the tweezer tips onto the sample box surface next to the chip, using the resulting torque spike to detect contact.
Subsequently, the tweezer tips are raised in steps of $\qty{0.1}{\milli\meter}$ until the change in torque between steps decreases below \qty{0.8}{\percent}. 
This signals that the tips have lifted off the bottom surface of the sample box. 
From this we obtain the height of the surface on which the chips rest and choose to grasp the chip \qty{0.2}{\milli\meter} above this surface.
A typical torque change versus step height is shown in Fig.~\ref{fig:fab}(c). 
Note that the torque does not change as much anymore once the step height goes above a specific value which means that the tweezers are now released from the surface.

The gripper then proceeds to close by a predefined amount to grasp the chip. The slightly compliant tweezer arms create a solid grip as the robot moves the chip to a beaker of developer solution (IPA:MIBK 3:1). The robotic arm swirls the chip in the solution in a circular motion for a user-specified amount of time (\qty{40}{\second}). Afterwards, it moves to a second beaker filled with a rinse solution (IPA) and does the same swirling motion for $\qty{10}{\second}$. To dry the chip, the robotic arm holds the chip in front of a USB-controlled $\text{N}_2$ gun on top of a cleanroom wipe to absorb the solvent. We find that for our configuration, drying for \qty{10}{\second} leaves the chip free of drying-induced residue. Finally, the robot returns the chip to its original position. It repeats the development for the remaining chips in the box and reuses the chip grasping height found for the first chip to save time. The full setup is shown in Fig.~\ref{fig:fab}(d) where the ChArUco calibration board is in the back right and the blow-drying on the left.   

To investigate whether robotic development presents significant advantages for process reliability, we choose to fabricate and characterize Josephson Junctions (JJ) where the e-beam resist development is done by either robotic or human operators. 
We will use JJ resistance variation as our primary metric, as it is a common benchmark for process optimization\cite{Kreikebaum.OBrien.ea:2020, Muthusubramanian.Finkel.ea:2024} and a direct indicator for the critical current of the JJ. 
The fabrication flow is designed to minimize differences between the chips developed by robotic and human operators, so that the most significant difference comes from developing itself. 
We start by forming a shallow grid with period $\qty{10}{\milli\meter}$ and $\qty{5}{\milli\meter}$ on the backside of a $4$-inch high resistivity ($>\qty[inter-unit-product = \cdot]{10}{\kilo\ohm \centi\meter}$) silicon wafer by partially dicing it to a depth of \qty{100}{\micro\meter}. 
After stripping the protective resist with acetone and isopropyl alcohol, we proceed to spinning the front side of the wafer with a bilayer of e-beam resist (\qty{270}{\nano\meter} PMMA on \qty{600}{\nano\meter} MMA) and doing e-beam lithography (Raith EBPG 5200+) on the wafer. 

The pattern consists of $52$ identical chips (see \ref{fig:wafer}(a)) and we take care to align the exposure with the backside grid. The pattern on each chip consists of $9$ rows of $12$ identical Josephson junction test devices, see Fig.~\ref{fig:fab}(e). The JJ electrode width is swept from \qty{50}{\nano\meter} to \qty{350}{\nano\meter} over the rows. The junction geometry is shown in \ref{fig:resistance}(a). After the exposure, and before developing, we cleave the wafer into individual chips which is made easier thanks to the pre-diced grid. By cleaving the wafer into chips, we can develop each of them separately. Among all the chips, we select $18$ of them based on the quality of the cleaved edges and distribute the chips into two groups: robot ($9$ chips) and human operator ($9$ chips) groups. The human group is further subdivided into three groups with three chips each, and where each group is assigned to a different experienced human operator (Human A, B and C). The $9$ chips in the robot group are then developed by the robotic arm while in parallel, the human operators develop $3$ chips each. 
Because of the jagged edges from imperfect cleaving, the robot lost grip on $3$ of the samples during development. We replaced the dropped samples and they were subsequently successfully developed by the robot. 

The next step is deposition of the aluminum/aluminum oxide JJs in an evaporator (Plassys MEB550S). To load all the developed samples simultaneously into the chamber, we first glue them with PMMA on a carrier wafer on which a grid has been pre-diced (see \ref{fig:wafer}(b)). The grid is important because the aluminum  evaporation for the JJ is angled and therefore the orientation matters. After the aluminum evaporation, the liftoff is done in a \qty{80}{\degreeCelsius} solution of NMP, thereby completing the fabrication.\cite{Kelly:2015} A microscope image of a single device is shown in Fig.~\ref{fig:fab}(f).

\begin{figure}[!htbp]
\includegraphics[scale=0.94]{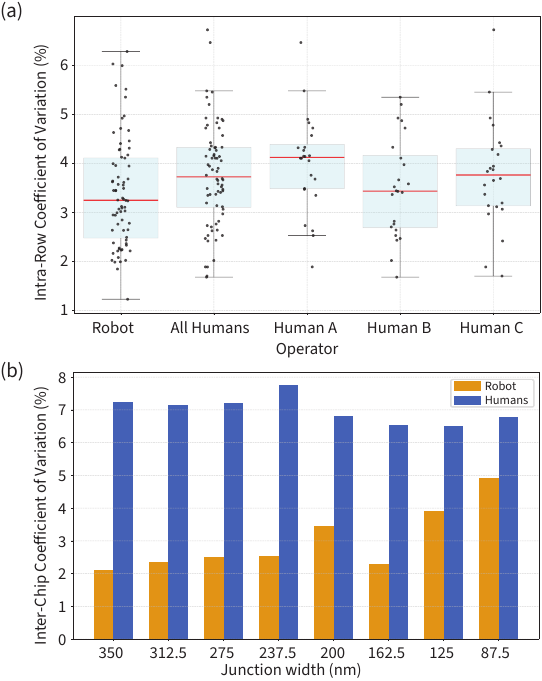}
\caption{\label{fig:results} \textbf{Robotic vs. human fabrication consistency.} 
\textbf{(a)}\,Intra-chip consistency of junction resistance. Box plot of the intra-row coefficient of variation (CV) for all device rows, grouped by operator. The red line is the median, the blue box spans the interquartile range (IQR, 25th to 75th percentile) and the whiskers extend to $ 1.5 \times$ IQR. Individual data points are overlaid as black dots. The robot group exhibits a lower median CV compared to the 'All Humans' group. 
\textbf{(b)}\,Inter-chip consistency of junction resistance. Bar chart comparing the CV for the robot (orange) and 'All Humans' (blue) groups for different junction widths. The CV is computed from the set of 9 mean resistances (one from each chip) per operator group for a given junction width.}
\end{figure}

We proceed to characterizing the resistance of the JJs at room temperature. The resistance is related to JJ critical current\cite{Ambegaokar.Baratoff:1963} and therefore the superconducting qubit frequency. Since the number of devices across $18$ chips totals $1944$, we use an automatic probe station (Micromanipulator P200L) to measure the IV curves, and extract the resistance of each JJ. The device yield is not \qty{100}{\percent} and before analyzing the results, we discard non-functional devices (shorted and open) and statistical outliers. 
First, we discard the entire row of \qty{50}{\nano\meter} wide JJs due to low yield, and difficulty in distinguishing a working JJ from an open device, which removes 216 devices. 
Next, to systematically remove the statistical outliers, we apply a row-wise $1.5\times\text{IQR}$ filter. For each junction geometry (row), we calculate the first and third quartiles (Q1, Q3) across all chips and remove measurements falling outside the range [$\text{Q1} - 1.5\times\text{IQR}$, $\text{Q3} + 1.5\times\text{IQR}$]. This step leads to the removal of 118 outliers. Finally, 9 devices were damaged by glue or a scratch during the fabrication process, leaving a total of 1601 high-quality measurements for the final analysis.

Our objective is to quantify the difference between the robot and human operator groups. We start by looking at the spread of resistances across identical devices on the same chip. For each row on every chip, we compute the intra-row coefficient of variation (CV) of the resistances, defined as the ratio of the standard deviation $\sigma$ to the mean resistance $\mu$ for that particular row, i.e. $\text{CV} = \sigma / \mu$. Calculating the CV instead of just the standard deviation allows us to normalize the spread, so that we can compare results across chips and geometries, even for groups of devices with different mean resistances. A larger intra-row CV indicates more disorder across identical devices on the same chip. We plot the intra-row CV for every row as black points in Figure~\ref{fig:results}(a) and separate the points into the different operators. The data points are overlaid on a box plot that summarizes the distribution of the CVs. For the robot, we obtain a median intra-row CV of \qty{3.1}{\percent} compared to \qty{3.6}{\percent} for the 'All Humans' group, in line with usual fabrication uncertainties\cite{Pop.Fournier.ea:2012, Kreikebaum.OBrien.ea:2020}. This shows that with the robot, identical devices on the same chip have an equal or slightly lower spread in resistance compared to a chip developed by a human operator.

While the previous metric looks at consistency within a chip, we can also analyze how resistance values vary across chips for nominally identical devices. The goal is to see if there are any variations between operators for chip development. To do this, we compute the mean resistance for each geometry on each chip and separate it into a robot and human group. These values are displayed in \ref{fig:resistance}(b) and (c). For each geometry, we take the standard deviation of the $9$ mean resistances and divide by the mean of the means. This gives us the inter-chip CV for each row which describes the spread in mean resistance across different chips. A low inter-chip value means that the mean resistance does not vary much from chip to chip. We plot the results as a bar chart in Fig.~\ref{fig:results}(b). For the wider JJs, we see that the CV is about $2$-$3$ times lower for the robot than for the human operator group. For example, for the \qty{350}{\nano\meter} wide JJs, we measure a \qty{2.1}{\percent} spread in mean resistance between the robot chips, compared to \qty{7.2}{\percent} between the human chips. For narrower JJs, the difference becomes smaller, possibly due to other process variations becoming more dominant. This variability is accentuated by the involvement of three different human operators, highlighting how individual stirring techniques and timing introduce significant disorder. While a single expert human might achieve high consistency, the robotic system ensures this baseline performance is maintained regardless of the operator, effectively removing the 'human factor' from the process equation.

In conclusion, we have deployed a robotic arm for automatic chip-scale resist development 
that integrates camera-based chip detection, torque-feedback height calibration and a tweezer toolhead.
Initial tests show that this system provides superior consistency over pooled human operators, demonstrating that robotic automation effectively standardizes the process across different users. 
This is particularly valuable in academic settings with high personnel turnover,
and our results effectively highlight the process disorder generated by different human operators.

We expect devices with small critical dimensions, such as superconducting qubits or photonic crystals to benefit most from this improved reliability.
As highlighted in Fig.~\ref{fig:concept}, other process steps are also likely to see an improvement in reliability and consistency by switching to robotic operation. For instance, the automated and precise gluing of small chips onto carrier wafers would ensure consistent location and orientation, improving consistency during subsequent etching or deposition steps in tools designed for full wafers.
The increased reliability, which reduces inter-chip spread to levels comparable with intra-chip spread, opens up explorations of processes that rely on systematic variations between chips, such as testing different development times, deposition conditions, or etching conditions.
Furthermore, automating such processes significantly reduces human exposure to hazardous cleanroom chemicals, such as Piranha (sulfuric acid and hydrogen peroxide), warm hydrofluoric acid\cite{Hartung.Kley.ea:2008} or the developer tetramethylammonium hydroxide (TMAH)\cite{Lee.Wang.ea:2011}.

For this system to be truly compatible with the flexible nature of a research environment, it must be adaptable.
A cleanroom user testing a minor process variation should not need to reprogram the robot's entire linear sequence.
Future work will therefore focus on integrating a large language model (LLM) to dynamically generate or modify process scripts based on simple user requests.
Furthermore, adding a real-time supervision model to detect and correct manipulation errors, such as a misplaced chip, will be crucial for robust, long-term deployment.

Looking forward, as embodied artificial intelligence advances, more complex and high-stakes tasks are becoming increasingly feasible. 
This includes the precise, rotationally-aligned stacking of 2D materials for van der Waals heterostructures, the hybrid micro-assembly of disparate components, like photonic lasers and silicon chips, or the high-precision alignment of optical fiber arrays.
All of these can be performed inside an inert atmosphere to reduce surface impurities.
Automating these delicate tasks, which are notoriously difficult for humans, promises to vastly improve process control and unlock new research avenues across physics and materials science.

\section*{Acknowledgments}
The authors would like to thank Kaveh Pezeshki, Samuel Engel, Nelson Ooi, Hubert Stokowski and Matthew~P. Maksymowych for fabrication assistance and for useful discussions.
This work was primarily funded by Amazon Web Services Inc.\ and US federal government via the US Department of Energy through grant no.\ DE-AC02-76SF00515 (Q-NEXT Center) and the U.S. Army Research Office (ARO)/Laboratory for Physical Sciences (LPS) Modular Quantum Gates (ModQ) program (Grant No. W911NF-23-1-0254). 
We are grateful for support for this work from the
Gordon and Betty Moore Foundation, Grant 12214.
S.G. acknowledges support by the Swiss National Science Foundation (SNSF) [225443] and the Knut and Alice Wallenberg foundation [KAW 2021-0341].
Part of this work was performed at nano@stanford RRID:SCR\_026695.

Fig.~\ref{fig:concept} was illustrated with the help of generative image models. 
The text was edited for clarity, grammar, and style using generative multi-modal models.

\section*{Competing interests}
A.H.S.-N. is an Amazon Scholar.

\section*{Data availability}
An overview of the project can be found on \href{https://www.youtube.com/watch?v=hRR8uEAh57A}{YouTube} (\url{https://www.youtube.com/watch?v=hRR8uEAh57A}).
A video of the developing of the devices, including the robotic arm and the three human operators, presented here can be found on \href{https://youtu.be/mcq2ElSzgdM}{YouTube} (\url{https://youtu.be/mcq2ElSzgdM}).
The code is released on GitHub \href{https://github.com/alps-ml/paper.robotic-resist-development}{alps-ml/paper.robotic-resist-development} (\url{https://github.com/alps-ml/paper.robotic-resist-development}).
A version of the code, as well as the raw data used to generate the figures is available at \href{https://zenodo.org/doi/10.5281/zenodo.17636138}{Zenodo} \url{https://zenodo.org/doi/10.5281/zenodo.17636138}.

\bibliography{ref}%

\clearpage
\newpage
\onecolumngrid

\section*{Supplementary information}

\renewcommand{\figurename}{}
\setcounter{section}{0}
\renewcommand{\thesection}{Supplementary Note \arabic{section}}
\renewcommand{\theHsection}{Sn\arabic{section}}

\setcounter{figure}{0}
\renewcommand{\thefigure}{Supplementary Figure \arabic{figure}}
\renewcommand{\theHfigure}{S\arabic{figure}}

\setcounter{table}{0}
\renewcommand{\tablename}{Supplementary Data Table}
\renewcommand{\theHtable}{S\arabic{table}} %

\section{Other sources of process variation}
\begin{figure}[!htb]
\includegraphics[scale=1.0]{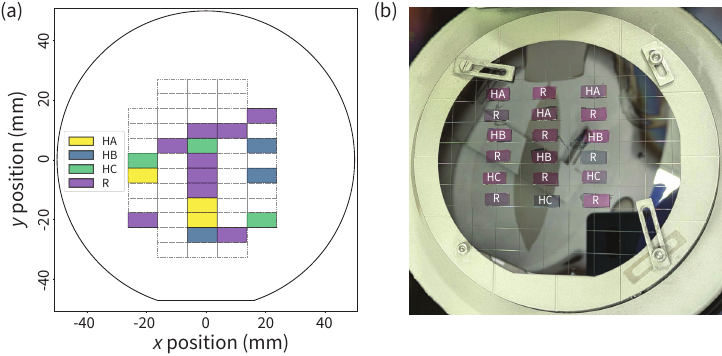}
\caption{\label{fig:wafer} \textbf{Chip location during fabrication.} 
\textbf{(a)}\,Location of the chips on the original $4$-in wafer during e-beam lithography, sorted by operator during developing. HA (B, C) stands for Human A (B, C) and R for Robot. 
\textbf{(b)}\,Chips glued to a carrier wafer for the aluminum deposition step. The carrier wafer has a pre-diced grid to facilitate alignment of the chips to the tilting axis during evaporation.}
\end{figure}

Apart from developing, we can identify a few other reasons for process uncertainty.
The location of the chip during e-beam lithography and during deposition can have an effect on process variation. 
In \ref{fig:wafer}(a), we show the location of the chips in the robot and human operator groups before cleaving. 
Since resist thickness is not perfectly uniform, this can influence the effective overlap in the junction area, and therefore could lead to some process variation.
We report the resist thicknesses for every chip in Supplementary Data Table~\ref{tab:resist_thickness_data}. 
We find that the resist thickness spread is similar for both the robot and human operator groups.
We do not expect this to be a large influence in our case, evident also by the fact that our resistance spread is comparable to other work in the field \cite{Pop.Fournier.ea:2012}.
Another possible source of process variation between chips could arise from the location of the chips in the deposition chamber which would affect film thickness. 
We distribute the chips on the carrier wafer by interleaving chips developed by humans and the robot to cancel out this effect as much as possible.

\begin{table*}[!hbp]
\centering
\caption{\label{tab:resist_thickness_data} \textbf{Thickness measurements of PMMA and MMA layers} using Reflectometry (Filmetrics F40). 
The thickness at the chip center is interpolated based on a \qtyproduct{10 x 10}{\milli\meter} grid.}

\sisetup{round-mode=places, round-precision=0, table-format=3.1}

\begin{minipage}[t]{0.48\textwidth}
    \vspace{0pt}
    \centering
    \begin{tabular}{@{}lc S[table-format=3.1] S[table-format=3.1]@{}}
    \toprule
    \textbf{Operator} & \textbf{Replicate} & {PMMA (\unit{\nano\meter})} & {MMA (\unit{\nano\meter})} \\
    \midrule
    HA & 1 & 261.0 & 596.0 \\
       & 2 & 263.0 & 600.0 \\
       & 3 & 261.0 & 592.0 \\
    \midrule[0.5pt] %
       & \textit{Mean} & 261.7 & 596.0 \\
       & \textit{Std. Dev.} & 1.2 & 4.0 \\
    \midrule
    HB & 1 & 262.0 & 594.0 \\
       & 2 & 264.0 & 597.0 \\
       & 3 & 261.0 & 592.0 \\
    \midrule[0.5pt]
       & \textit{Mean} & 262.3 & 594.3 \\
       & \textit{Std. Dev.} & 1.5 & 2.5 \\
    \midrule
    HC & 1 & 262.0 & 598.0 \\
       & 2 & 288.0 & 614.0 \\
       & 3 & 261.0 & 592.0 \\
    \midrule[0.5pt]
       & \textit{Mean} & 270.3 & 601.3 \\
       & \textit{Std. Dev.} & 15.3 & 11.4 \\
    \midrule
    \multicolumn{2}{l}{\textbf{Mean (Humans)}} & 264.8 & 597.2 \\
    \multicolumn{2}{l}{\textbf{Std. Dev. (Humans)}} & 8.8 & 6.9 \\
    \bottomrule
    \end{tabular}
\end{minipage}%
\hfill
\begin{minipage}[t]{0.48\textwidth}
    \vspace{0pt}
    \centering
    \begin{tabular}{@{}lc S[table-format=3.1] S[table-format=3.1]@{}}
    \toprule
    \textbf{Operator} & \textbf{Replicate} & {PMMA (\unit{\nano\meter})} & {MMA (\unit{\nano\meter})} \\
    \midrule
    \multirow{9}{*}{Robot} 
     & 1 & 263.0 & 598.0 \\
     & 2 & 261.0 & 592.0 \\
     & 3 & 266.0 & 610.0 \\
     & 4 & 281.0 & 608.0 \\
     & 5 & 290.0 & 615.0 \\
     & 6 & 275.0 & 612.0 \\
     & 7 & 261.0 & 590.0 \\
     & 8 & 278.0 & 613.0 \\
     & 9 & 286.0 & 613.0 \\
    \midrule[0.5pt]
    \multicolumn{2}{l}{\textbf{Mean (Robot)}}      & 273.4 & 605.7 \\
    \multicolumn{2}{l}{\textbf{Std. Dev. (Robot)}} & 11.1  & 9.7 \\
    \bottomrule
    \end{tabular}
\end{minipage}
\end{table*}

\section{Mean Josephson junction resistance values}
\label{SI:resistance}
\begin{figure}[!htb]
\includegraphics[scale=1.0]{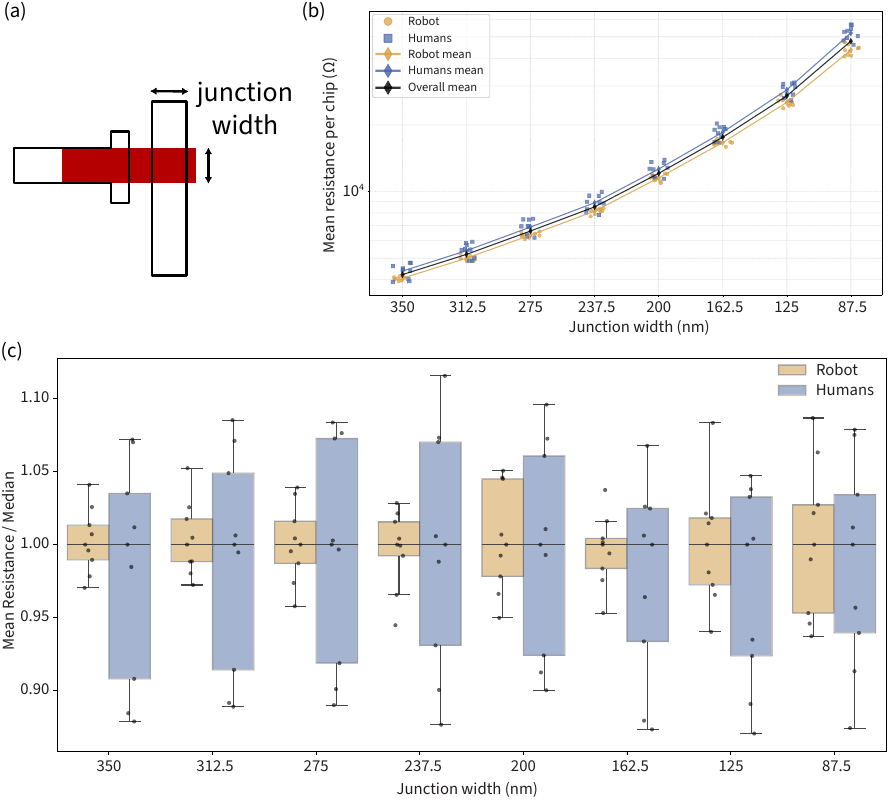}
\caption{\label{fig:resistance} \textbf{Josephson junction resistance.} 
\textbf{(a)}\,Josephson junction design, showing what we refer to as JJ width. 
\textbf{(b)}\,Mean JJ resistance vs JJ width for the 18 chips, separated into robot (orange) and human (blue) operators. Mean over all the robot (human) chips are marked with an orange (blue) diamond, and the overall mean with a black diamond marker. 
\textbf{(c)}\,Mean resistance on each chip normalized by the respective median, and grouped by junction width and operator (black data points). A box plot is overlaid on top of the points showing that the robot group has a smaller resistance spread across chips than the 'All humans' group. The horizontal black line in the box plot is the median.}
\end{figure}

In Fig.~\ref{fig:results}(b), we showed the inter-chip CV for each JJ width (the JJ geometry is shown in \ref{fig:resistance}(a)). This CV was calculated based on the mean JJ resistance on each chip for a particular JJ geometry. We plot these mean resistance values in \ref{fig:resistance}(b). 
From the data points, we can see that the human operator resistance values are more spread out compared to the robot ones. 
The CV we compute in the main text quantifies this spread. To better visualize this spread, we normalize these resistance values by the respective median and plot them overlaid on a box plot in \ref{fig:resistance}(c), showing that while an individual human operator might be as consistent as a robotic arm, there is a larger spread among human operators.

\section{Hand-eye calibration}
\label{SI:hand}

\begin{figure}[!htb]
\includegraphics[scale=1.0]{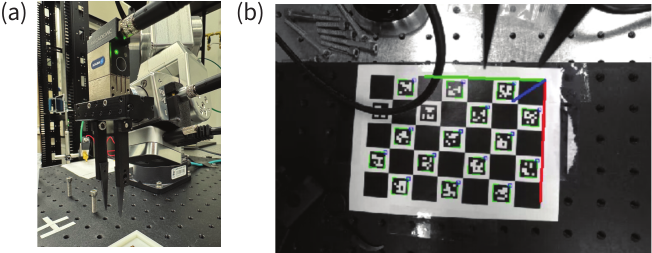}
\caption{\label{fig:handeye} \textbf{Hand-eye calibration.} 
\textbf{(a)}\,Close-up of the robot end-effector and camera. 
\textbf{(b)}\,ChArUco board overlaid with marker detection boxes, found using OpenCV via the MoveIt2 calibration package.}
\end{figure}

The camera is mounted to link 5 of the Mecademic robot using a custom mount, see \ref{fig:handeye}(a). A common challenge in robotics is to map the position and orientation of objects from the camera frame to the end-effector frame. When we detect chips using the camera, we need to transform the chip position in the camera frame to the position in the Tool Center Point (TCP) frame which is at the tweezer tips (i.e. the end-effector). The tweezers themselves are mounted to link $6$ of the robot. While the transformation from link $6$ to the tweezer tips is a simple offset, the transformation from the camera frame to link $6$ is more involved. We construct a Unified Robot Description Format (URDF) of the Mecademic robot which allows us to do the transformation from link $5$ to link $6$ using the tf2 library~\cite{foote2013tf}. For the camera frame to link $5$, we use the MoveIt2 calibration package~\cite{orsula2024moveit2calibration} to compute this transformation based on the Park~1994 method~\cite{park1994robot}. To do this, we take $9$ sample pictures of a $5\times7$ ChArUco board (see \ref{fig:handeye}(b)) with the camera mounted to the robot, making sure to include a diverse set of joint states for the robot. We use a board with marker size of \qty{14.2}{\milli\meter} and a total board length of \qty{148.1}{\milli\meter}, and the DICT\_5X5\_250 ArUco dictionary from OpenCV. We don't perform any additional calibration on the RealSense camera which is precalibrated by the manufacturer. 

One challenge is that the chips are very thin (\qty{0.5}{\milli\meter}), making it hard to accurately measure the correct grasping height using the camera. When applying the transformations, we therefore assume that the chip is parallel to the table and only take the in-plane ($x$-$y$) coordinates and the in-plane rotation into account. For finding the height ($z$), we use the much more precise torque feedback as explained in the main text. 

The resulting set of transformations corresponds to our hand-eye calibration. We are able to accurately grab a $\qtyproduct{10 x 5}{\milli\meter}$ chip on its short edge thanks to our calibration being accurate within $\sim \qty{3}{\milli\meter}$. We did not observe any deterioration in the calibration accuracy over a period of $4$ months, showing that our camera mount is robust.

\end{document}